\documentclass[letter,twocolumn]{emulateapj}
\usepackage{amstext}
\usepackage{apjfonts}
\usepackage{amsmath}
\usepackage{amssymb}
\usepackage{graphicx}

\usepackage[usenames,dvipsnames]{color}

\newcommand{\SNII}{SN II}
\newcommand{\SNIa}{SN Ia}

\shorttitle{\textsc{R-Process Enrichment in the Milky Way}}
\shortauthors{\textsc{Shen et al.}}

\begin{document}


\newcommand{\msun}{{\rm M}_{\odot}}

\title{The History of R-Process Enrichment  in the Milky Way}

\author{Sijing Shen\altaffilmark{1}, Ryan J. Cooke\altaffilmark{2}, Enrico Ramirez-Ruiz\altaffilmark{2}, Piero Madau\altaffilmark{2}, Lucio Mayer\altaffilmark{3} and Javiera Guedes\altaffilmark{4}}

\altaffiltext{1}{Institute of Astronomy, University of Cambridge, Madingley Road, Cambridge, CB3 0HA, United Kingdom}
\altaffiltext{2}{Department of Astronomy and Astrophysics, University of California, 1156 High Street, Santa Cruz, CA 95064, USA}
\altaffiltext{3}{Institute of Theoretical Physics, University of Z\"{u}rich, Winterthurerstrasse 190, CH-9057 Z\"{u}rich, Switzerland}
\altaffiltext{4}{Teralytics AG, Zollstrasse 62, 8005 Z\"{u}rich, Switzerland}

\begin{abstract} 

We investigate the production sites and the enrichment history of $r$-process elements in the Galaxy, as traced
by the [Eu/Fe] ratio, using the high resolution, cosmological zoom-in simulation `Eris'. At $z=0$, Eris represents
a close analog to the Milky Way, making it the ideal laboratory to understand the chemical evolution of our Galaxy.
Eris formally traces the production of oxygen and iron due to Type-Ia and Type-II supernovae.
We include in post-processing the production of $r$-process elements from compact binary
mergers. Unlike previous studies, we find that the
nucleosynthetic products from compact binary mergers can be incorporated into stars of very low metallicity and
at early times, even with a minimum delay time of 100 Myr. This conclusion is relatively insensitive to modest variations
in the merger rate, minimum delay time, and the delay time distribution. By implementing a first-order prescription for metal-mixing, we can
further improve the agreement between our model and the data for the chemical evolution of both [$\alpha$/Fe] and [Eu/Fe].
We argue that compact binary mergers could be the dominant source of $r$-process nucleosynthesis in the Galaxy.

\end{abstract}


\section{Introduction}

The chemical abundance patterns of Galactic halo stars encode precious information about the various stellar progenitor systems that existed prior to their birth. These ancient halo stars therefore provide an insight into the nucleosynthesis processes that occurred early in the history of the Milky Way.  Of particular interest in this regard is the heavy element composition of Galactic halo stars \citep{truran2002,cowan2006}. 
For stars with an [Fe/H] metallicity in the range $\approx10^{-2}$ to $\approx10^{-3}$ solar, elements in the mass region above Ba have been found to be consistent with enrichment by a pure $r$-process with a distribution that is characteristic of solar system matter but with a large star-to-star bulk scatter in their concentrations with respect to the lighter elements such as Mg \citep{Sne08}.
The presence of these heavy nuclei in such primitive stars demonstrates that the $r$-process has operated in a fairly robust manner over large periods of time in Galactic history, while the large dispersion in their abundance relative to lighter nuclei suggests an early, chemically unmixed and inhomogeneous Milky Way \citep{fields2002}. At later times, these localized inhomogeneities would be smoothed out as subsequent events take place and heavy element  products are given more time to migrate throughout the Galaxy \citep{Tra2001}.

Nucleosynthesis theory has identified the specific physical conditions and nuclear properties required for the $r$-process \citep{Bu1957}. However, the astrophysical site for this process has not been unambiguously identified.
The original work on this subject suggested that the neutron-rich regions outside a nascent neutron star in a Type II supernova \citep{woosley1994,takahashi1994} or the ejecta from the last seconds of a merger between a neutron star (NS) and a compact binary companion are the most likely formation sites \citep{Lat1977,Fre1999}. Compact mergers involving a neutron star are much rarer than \SNII\ \citep{cowan2004} and should occur far from their birth sites \citep[e.g.][]{Ke2010}. Furthermore, these two mechanisms eject different quantities of $r$-process material. These differences should surely be imprinted in the enrichment pattern of $r$-process elements and may ultimately identify the dominant production mechanism \citep{Arg04,Mat14,Cescutti13,TsuShi14,CesChi14}.

The build-up of the elements in our Galaxy, including that of the $r$-process, can be studied in detail by using realistic galactic chemical evolution (GCE) models (e.g. \citealt{Pag09}). 3D hydrodynamic simulations that incorporate the details of chemical evolution (often referred to as chemodynamical simulations) have been widely used in the literature to study the enrichment history and distribution of various elements in galaxies \citep[e.g.][]{Kawata03,KobNak11,Rahimi11,Few12,Pilkington12,Minchev13,Brook14,Few14}. Because such simulations follow the dynamics and chemical evolution self-consistently, they are better equipped to address the inhomogeneous enrichment of the ISM and the mixing of metals. In addition, simulations performed in a cosmological context \citep[e.g.,][]{KobNak11,Rahimi11,Pilkington12,Brook14,Few14} are able to capture the larger scale mixing mechanisms including gas inflows, satellite mergers, galactic winds and fountains, and instabilities in rotationally supported disks, all of which are undoubtedly important to understand the enrichment and dispersal of heavy elements. On the other hand, since cosmological simulations today are limited by resolutions of a few tens to hundreds of parsecs, they inevitably involve ``sub-grid'' models for star formation, stellar feedback, and/or turbulent mixing. As such, large uncertainties still exist at the hundred parsec scale. Investigating the distribution of chemical elements in these simulations and comparing them extensively with the observational data can, in turn, provide important constraints on these sub-resolution models and improve the modeling of astrophysical fluids.

The origin and evolution of the $r$-process elements, however, are mostly addressed using analytical or semi-analytical models (e.g. \citealt{Arg04,Mat14,Cescutti13,TsuShi14,CesChi14}; although see \citealt{vdVoort15}), possibly because most simulations have not implemented the production of $r$-process elements due to their uncertain origins described above. In this Paper we employ Eris, one of the highest resolution cosmological simulations of the formation of a Milky Way-size galaxy \citep{Guedes11}, to investigate the synthesis of the heavy $r$-process elements in our Galaxy, as traced by the [Eu/Fe] ratio.

\section{Methods}

\subsection{The Eris Simulation}
\label{sec:erissim}
We use the high resolution, zoom-in cosmological simulation of a Milky Way Galaxy analog ``Eris'' to track the production and transportation of $r$-process elements. A detailed description of the Eris simulation is provided by \citet{Guedes11}. Here we briefly outline the aspects relevant to this study. The simulation was performed with the parallel TreeSPH code \textsc{Gasoline} \citep{Wadsley04} in a {\it WMAP-3} cosmology.
The run includes a uniform UV background \citep{HM96}, Compton cooling, atomic cooling and metallicity dependent radiative cooling at T $<10^{4}$ K. Star formation is modelled by stochastically forming ``star particles'' out of gas that is sufficiently cold ($T < 3 \times 10^{4}$ K) and reaches a threshold density of $n_{\rm SF}=5$ atoms cm$^{-3}$. The local star formation rate follows $d\rho_*/dt=0.1 \rho_{\rm gas}/t_{\rm dyn} \propto \rho_{\rm gas}^{1.5}$, where $\rho_*$ and $\rho_{\rm gas}$ are the stellar and gas densities, respectively, and $t_{\rm dyn}$ is the local dynamical time. Each star particle has initial mass $m_{\ast} = 6000 \  \msun$ and represents a simple stellar population that follows a \citet{Kroupa93} initial mass function (IMF), and inherits the metallicity of its parent gas particle. Star particles inject energy, mass and metals back into the ISM through Type Ia, Type II SNe and stellar winds \citep{Stinson06}. Eris' high resolution enables the development of an inhomogeneous ISM which allows realistic clustered star formation and strong cumulative feedback from coeval supernova explosions. Large scale galactic winds are launched as a consequence of stellar feedback, which transports a substantial quantity of metals into the circumgalactic medium and enriches the subsequent gas accretion \citep{Shen13}. At $z = 0$, Eris forms an extended, rotationally supported stellar disk with a small bulge-to-disk ratio. The structural properties, the mass budget in various components and the scaling relations in Eris are simultaneously consistent with observations of the Galaxy \citep{Guedes11}. 

The simulation follows \citet{Raiteri96} to model metal enrichment from SN II and SN Ia. Metals are distributed to gas within the SPH smoothing kernel (which consists of 32 neighboring particles). For SN II, metals are
released as the main sequence progenitors die, and iron and oxygen are produced according to the following fits to the \citet{Woosley95} yields:   
\begin{equation}                                                                
M_{\rm Fe} = 2.802 \times 10^{-4} \left({m_*\over \msun}\right)^{1.864}\,\msun,                                   
\end{equation}                                                                  
and
\begin{equation}                                                                
M_{\rm O} = 4.586 \times 10^{-4} \left({m_*\over \msun}\right)^{2.721}\,\msun. 
\end{equation}         
                                                                                                                  
For \SNIa, each explosion produces 0.63 $\msun$ of iron and 0.13 $\msun$ of oxygen \citep{Thielemann86}. 
Stellar wind feedback is based on \citet{Kennicutt94}, and the returned mass fraction was determined following \citet{Weidemann87}. The returned gas inherits the metallicity of the star particle. We adopt the \citet{Asplund09} solar abundance scale for elements other than O and Fe, that are not tracked in the simulation. 

We note that there are several limitations to our modeling. First, because the smallest gas resolution element is an ensemble of gas particles within the smoothing kernel rather than individual particles, forcing the newly-formed star particle to inherit metallicity only from its parent gas particle may amplify sub-resolution metallicity variances between gas particles. Second, the simulation used a traditional SPH formalism where metals advect with the fluid perfectly, without mixing due to microscopic motions (such as turbulence). Both caveats may cause an artificially inhomogeneous chemical distribution \citep{Wie09,SheWadSti10}. While improved runs within the Eris suite include a model for turbulent diffusion \citep{Shen13}, for this study we have used a simulation without mixing so that our $r$-process injection method is consistent with the production and distribution of O and Fe in the simulation. In Section~\ref{sec:results}, we explore a simple diffusion model to illustrate the effect of mixing. Instead of using the metallicity of the parent gas particle, we average the metallicity over 128 neighboring gas particles to determine the metallicity of the newly-formed star particle.

In addition, despite being one of the highest resolution galaxy formation simulations that evolves to z =0, our model still lacks the spatial and temporal resolution to correctly follow the formation and enrichment of the first generation of Population III stars as well as early Population II stars. It is a common practice in the literature to introduce a metallicity ``floor'' at high redshift to account for the unresolved earliest population \citep[e.g.,][]{Krumholz11,Kuhlen13,Hopkins14}. The floor metallicity is typically around $10^{-4}$ to $10^{-3}~\rm Z_{\odot}$, motivated by detailed simulations of Population III star formation and primordial metal enrichment \citep[e.g.][]{Wise12}. Although a metallicity floor was not included in Eris during the run, in Section~\ref{sec:results} we investigate the affect of a modest metallicity floor ($10^{-4}~Z_{\odot}$) implemented in post-processing. This is achieved by assigning a minimum abundance of [Fe/H]~$=-4.0$ to every gas particle, and an $\alpha$-enhancement of [O/Fe]~$+0.4$, corresponding to the IMF-weighted alpha-enhancement estimated from models of zero-metallicity stellar nucleosynthesis \citep{Woosley95,HegWoo10,LimChi12}.

\subsection{R-Process Production Sites and Injection History}
\label{sec:method}

The key ingredient in our analysis is the realistic
star formation history (SFH) of a galaxy simulated in a cosmological context which
at redshift $z=0$ is a close analog of the Milky Way \citep{Guedes11}.
In this section we describe
our post-processing implementation for NS mergers. 
The important elements of our model include: (1) The delay-time
distribution (DTD) of mergers; (2) the merger rate and the yield
of $r$-process elements; and (3) the spatial distribution of injection
sites and their sphere of influence.

\begin{figure}
\plotone{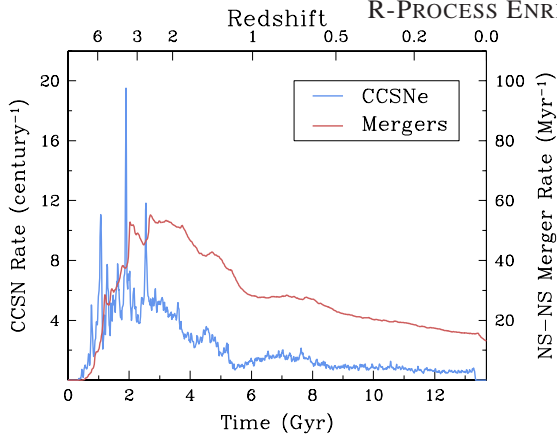}
\caption{The \SNII\ rate (blue curve, left axis), which closely traces the SFR, 
is compared with the NS merger rate derived from the Eris SFR and the DTD (red curve, right axis).}
\label{fig:sfr}
\end{figure}

\begin{figure*}
\plotone{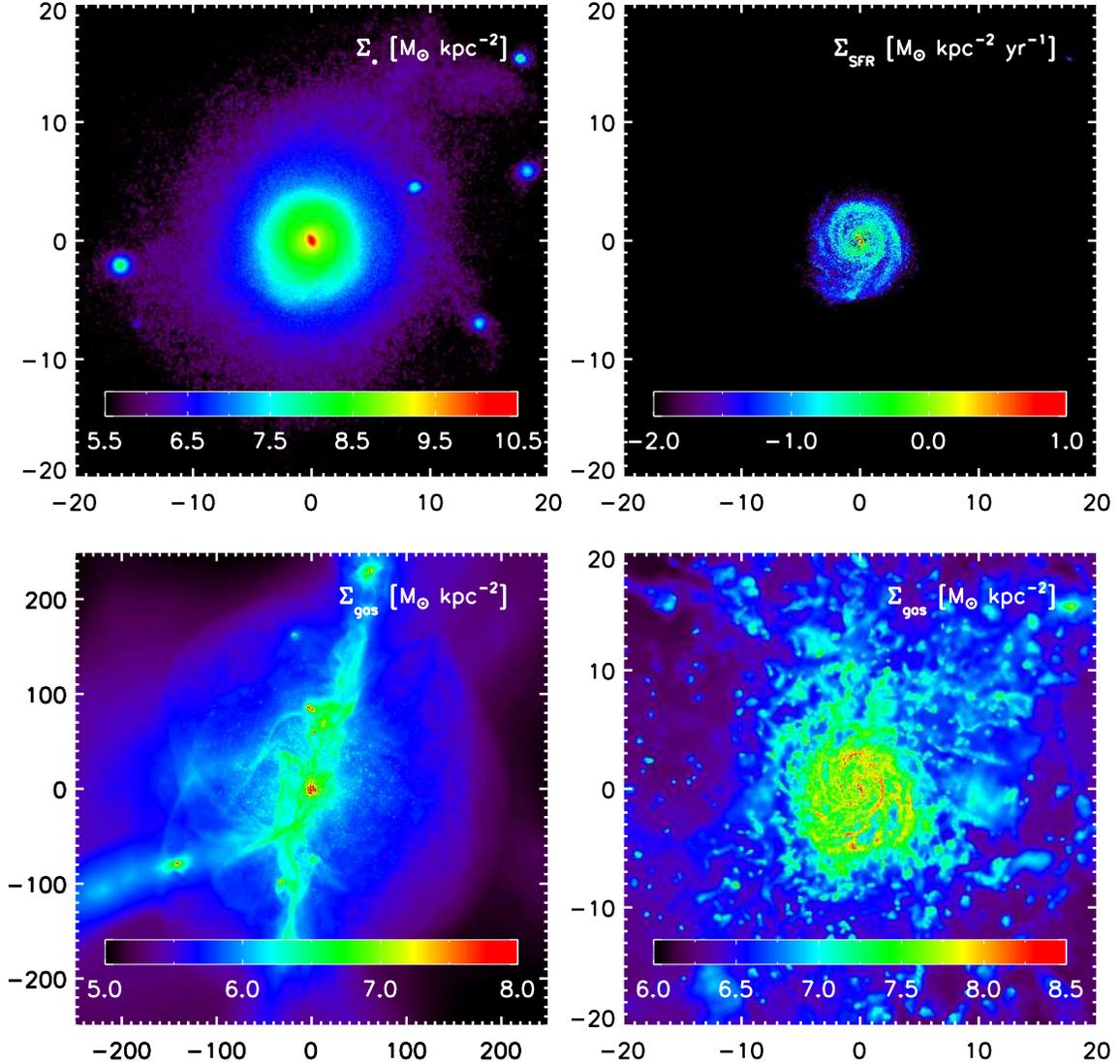}
\caption{A face-on illustration of the projected surface mass density of stars (top-left panel), SFR
surface density (top-right panel) and gas in different scales (bottom panels) for Eris at redshift $z=2$. The units of all axes are in kpc.
}
\label{fig:simplot}
\end{figure*}

\subsubsection{Delay-time distribution, merger rates and  r-process yields}

The merger DTD, $P(t)$, is well-modeled by  a power-law \citep{piran1992,Kal2001}, although there is significant uncertainty on the value of the exponent \citep[e.g.][and references therein]{Beh2014}. Herein, we adopt $P(t) \propto t^{-n}$ for $t > t_{\rm cut}$ and zero probability otherwise, where $t_{\rm cut}$ is the initial time delay after a burst of star formation before the first merger occurs. Herein we consider two power-law indices, $n = 2$ and $n = 1$, and assume conservative values of $t_{\rm cut}=100$ Myr and $ t_{\rm cut}=200$ Myr \citep{fryer1999,Bel06}. 

Our fiducial model assumes that each merger event produces a mass $M_{\rm rp} = 0.05~M_{\odot}$ of $r$-process elements \cite[e.g.][]{just2014}. We also consider a model where $M_{\rm rp} = 0.01~M_{\odot}$ is produced, which reflects the mass of dynamically ejected material \citep{Lattimer74,Rosswog99,Metzger10,Roberts11,Bauswein13,Grossman13,rr2014}.
For each merger event, we track the production of europium, which we assume is produced in solar relative proportions such that $M_{\rm Eu}/M_{\rm rp}=9.3\times10^{-4}$ \citep{Sne08}. 

The NS merger rate is then determined by convolving the SFH extracted from the simulation with the DTD:
\begin{equation}
\label{eq:mrate}
{\cal R}(t) = A \int^{t_{H}}_{t_{\rm cut}} \dot{M}_{\ast}(t-\tau)~P(\tau)~{\rm d}\tau  
\end{equation} 
where $A$ is a constant that is fixed by the total number of merger events,  $\dot{M}_{\ast}$ is the star formation rate (SFR) and $t_{H}$ is the Hubble time. To calculate $A$, we impose that the abundance of Eu/O in the final simulation output corresponds to the solar value (i.e. [Eu/O]~$= 0.0$). There are $\sim260$ million massive stars formed in Eris that end their life as a Type II SN, and the IMF-weighted O yield per Type II SN event is $\simeq 1 \ {\rm M}_{\odot}$ in our model. Assuming a solar number ratio, log(Eu/O)$_{\odot}=-8.14$, the total mass of Eu needed to obtain the solar [Eu/O] abundance is $\sim18~M_{\odot}$. Thus, the total mass of $r$-process elements produced during the chemical evolution of Eris needs to be $18\,800~{\rm M}_{\odot}$. If we now assume that each compact binary merger contributes $M_{\rm rp}=0.05~{\rm M}_{\odot}~(0.01~{\rm M}_{\odot})$ of $r$-process, then we require $\approx 3.76 \times 10^{5}~(1.88 \times 10^{6})$ mergers in 13.8 Gyr to explain the observed solar Eu/O ratio. The constant $A$ is therefore fixed by requiring that the integral of the merger rate (Eq.~\ref{eq:mrate}) over the lifetime of Eris is equal to $3.76 \times 10^{5}~(1.88 \times 10^{6})$. The resulting NS merger history is shown in Figure~\ref{fig:sfr} as the solid red curve, for $M_{\rm rp}=0.05~{\rm M}_\odot$, which is in good agreement with the expected rates calculated by \citet{Aba10}. For reference, we also present the Eris \SNII\ rate as a function of time as the blue curve in Figure~\ref{fig:sfr}.

\subsubsection{Injection History}

\begin{figure*}
\plotone{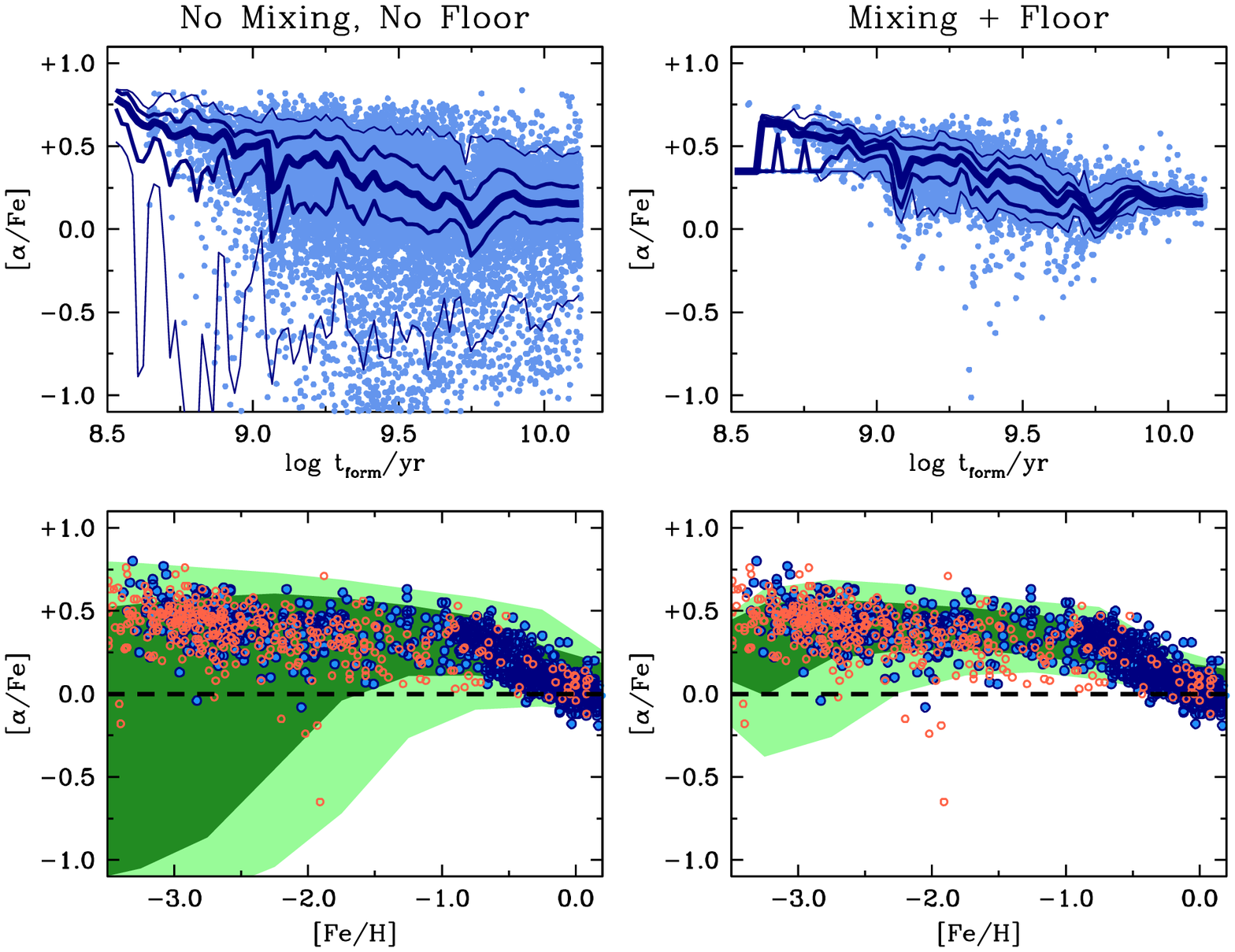}
\caption{\textit{Top panels}: The chemical evolution of O and Fe
for a representative sample (1:1000) of Eris' $z=0$ star particles (blue symbols) as a function of the formation time.
The centroid of the distribution is shown by the solid dark blue line, whilst the thinner lines
enclose 68 and 95 per cent of the stars.
The top-left displays the raw simulation output, without a metallicity floor and without a prescription for metal diffusion. The `excess'
scatter at low [$\alpha$/Fe] is reduced considerably when we include a
simple physical model for metal diffusion and a metallicity floor, as shown in the top-right panel (see text for further details).
\textit{Bottom panels}: The dark and light green contours show the [$\alpha$/Fe] distributions that
respectively enclose 68 and 95 per cent of Eris' $z=0$ star particles. The observational data are
represented by blue symbols for stars where $\alpha$, Fe, and Eu were all measured,
or by orange circles when only $\alpha$ and Fe are reported. The dashed line
represents the solar value. Our simple mixing prescription and introducing a
metallicity floor of $Z=10^{-4}\,Z_{\odot}$ significantly improves the agreement
with the observational data.
}
\label{fig:aFemix}
\end{figure*}

Compact binary mergers are expected to predominantly occur within
$\sim10-100$~kpc of a Milky Way-like host galaxy \citep{BloSigPol99,Bel06,Ke2010}.
Therefore, the spatial distribution of NS
mergers broadly follows the stellar distribution of the host
galaxy. We have thus adopted a post-processing implementation to
include NS mergers in Eris. Our 
approach is justified since the momentum imparted to the surrounding gas
by a merger is much less than that of a SN explosion, despite releasing
an energy that is similar in magnitude to a SN. The gas dynamics is 
therefore largely unchanged.

In the top-left panel of Figure~\ref{fig:simplot}, we present a face-on
illustration of the projected surface mass density of Eris' stars
at redshift $z=2$. By construction, this panel
represents the distribution of merger injection sites.
Similarly, in the top-right panel of this figure,
we present the SFR surface density, which traces
\SNII. In the bottom panels of Figure~\ref{fig:simplot},
we show the surface mass density of gas on two different
spatial scales, which are enriched by these events and
later form a new generation of stars.

Using the above formalism, we calculate the number of NS mergers that occur between adjacent timesteps and randomly select a corresponding number of star particles from a uniform distribution to act as the merger injection sites. The surrounding gas particles are then enriched with a total Eu mass $M_{\rm Eu}^{\rm tot}$~$=4.65\times10^{-5} \rm M_{\odot}$ (corresponding to $M_{\rm rp}$ = 0.05 $\rm M_{\odot}$), which is distributed over the 32 neighboring gas particles according to the smoothing kernel, as outlined in \citet{Wadsley04}.
We note that the oxygen and iron enrichment follow an identical scheme. The evolution of each gas particle is tracked for subsequent timesteps. When a stellar particle is born in the non-diffusion case, it inherits the Eu/Fe ratio from the parent gas particle. In our simple mixing model, the stellar particle inherits the average abundance of Eu, O and Fe from 128 neighbouring gas particles as described in Section \ref{sec:erissim}. Hereafter, when comparing with observations, we only consider star particles present in the $z=0$ snapshot.

\section{R-Process Enrichment in The Milky Way}
\label{sec:results}

\begin{figure*}
\plotone{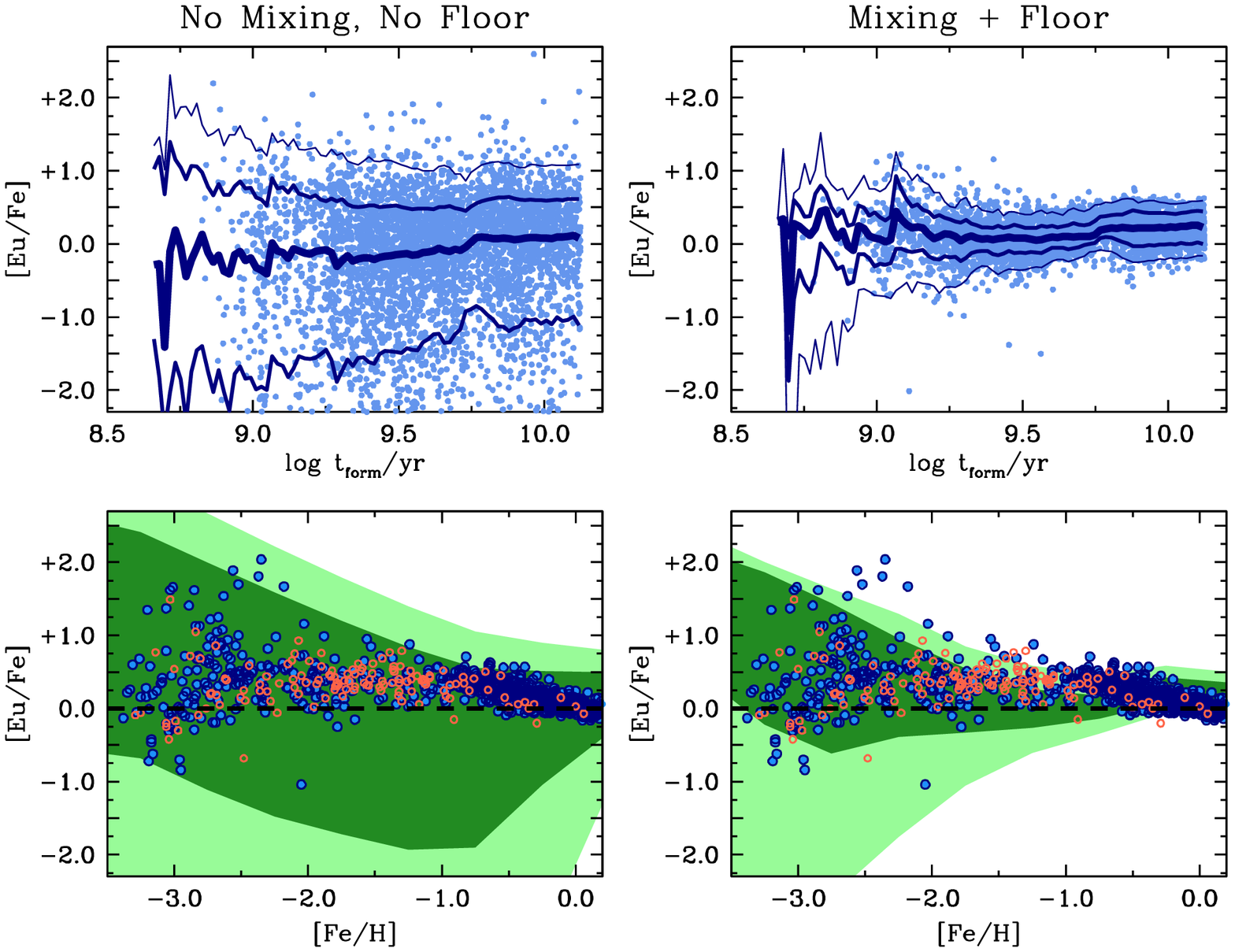}
\caption{
Same as Figure~\ref{fig:aFemix}, but illustrates the chemical evolution of [Eu/Fe] for our fiducial model,
which corresponds to a mass ejection of $M_{\rm rp}=0.05~{\rm M}_{\odot}$ per merger, a merger DTD that has a
power law form ($P(t)\propto t^{-1}$) and a minimum merger delay-time of $t_{\rm cut}=100$~Myr. In this figure,
orange symbols represent stars where only Eu and Fe are reported and blue symbols show stars where
$\alpha$, Eu and Fe are measured.}
\label{fig:EuFemix}
\end{figure*}

Our baseline set of model parameters include:
an $r$-process mass of $M_{\rm rp}=0.05~{\rm M}_{\odot}$ ejected per
event, a merger DTD that has a minimum merger
delay-time of $t_{\rm cut}=100$ Myr and a power law of the form $t^{-1}$,
a mixing length corresponding to the 128 nearest neighbors, and an
alpha-enhanced ([$\alpha$/Fe]=+0.40) metallicity floor of [Fe/H]~$=-4.0$.
Hereafter, we refer to this set of parameters as our fiducial model.

We now investigate the chemical evolution
of the $\alpha$ (traced by O), Fe and $r$-process elements in Eris.
The top-left panel of Figure~\ref{fig:aFemix} displays
the [$\alpha$/Fe] abundance for a representative sample
(1:1000) of Eris star particles in the $z=0$ snapshot as a function
of their formation time.
Although most stars exhibit super-solar [$\alpha$/Fe]
values, there are a non-negligible number of stars with
strongly sub-solar values. This effect is more pronounced
at the lowest metallicity, as shown in the bottom-left panel
of Figure~\ref{fig:aFemix}, where the distribution
of [$\alpha$/Fe] values is represented by dark and light green
contours that respectively enclose 68 and 95 per cent of the
Eris stars at a given metallicity.
When [Fe/H]~$\lesssim-1.5$, the [$\alpha$/Fe] distribution
bifurcates into a high and low [$\alpha$/Fe] channel. The low
[$\alpha$/Fe] channel corresponds to star particles
predominantly enriched by \SNIa\ (which, in our implementation,
produce an O mass that is five times less than the Fe mass), whereas
the high [$\alpha$/Fe] channel represents the star particles
enriched solely by \SNII.

To compare with observations, we overplot a sample of [$\alpha$/Fe]
measurements for Milky Way thin disk, thick disk, and halo stars
\citep{Ful00,Red03,Cay04,Coh04,Sim04,Ven04,Bar05,Reddy06,Mis13,Roe14}.
We divide the observational sample into stars where $\alpha$ (typically Mg, Si, Ca),
Fe, and Eu are \textit{all} measured (blue symbols), and stars where
only $\alpha$ and Fe are measured (orange symbols). Overall, there
is a reasonable agreement between the observations and the upper envelope
of the simulated stars. Moreover, the `knee' in the $\alpha$/Fe ratio near a
metallicity [Fe/H]~$\simeq-1.0$, which marks the increased
contribution of \SNIa\ \citep[e.g.][]{Tin79}, is well-reproduced
by the Eris simulation.

The discrepancy between Eris and the
observations at low metallicity is the result of gas
particles being predominantly enriched by \SNIa,
and illustrates a limitation of the traditional SPH formalism; 
once a gas particle is enriched with metals, it cannot
share its metals with neighboring particles. Many techniques
have been developed to incorporate metal diffusion
in simulations, using either subgrid turbulent diffusion models
\citep{Gre09,SheWadSti10}, or simply smoothing the
metals within the SPH kernal \citep{Wie09}. Our goal
is to provide a simple demonstration of the importance
of metal diffusion for studying chemical evolution. We
have therefore post-processed
Eris to include a metallicity floor (as described in Section~\ref{sec:erissim})
and we paint newly formed star particles with
a metallicity corresponding to the mass-weighted
average of the 128 nearest neighbors. 
The choice of 128 is based on the following estimation: At our star formation threshold density of 5 atoms cm$^{-3}$, this corresponds to a size that the gas can cross within the free-fall time, assuming the typical velocity dispersion of the molecular cloud. The corresponding mixing length of 128 particles peaks around $50-120~{\rm pc}$ at all redshifts. Only in very rare cases ($\sim2$ per cent) does the mixing length exceed $500~{\rm pc}$.  Even at high redshift (e.g. $z > 5$) where the main host progenitor undergoes vigorous mergers, less than 10 per cent of cases have a mixing length in excess of $350~{\rm pc}$. We have carried out a test run in which we remove all cases with smoothing length larger than 350 pc, and find that the results are essentially unchanged.
The chemical evolution of [$\alpha$/Fe] for our mixing prescription is presented in the right panels
of Fig.~\ref{fig:aFemix}, where the excess [$\alpha$/Fe] scatter
for the Eris star particles is considerably reduced at all times.
The data are brought into much better agreement with the
simulations at low metallicity. 
We emphasise that this simple model of metal diffusion is designed to illustrate the importance of subgrid mixing. Although our choice of the number of smoothing neighbours (or length) is physically motivated, 128 neighbours also provides the closest match to the observed alpha/Fe scatter seen in the simulations.  In other words, we have reduced the artificial inhomogeneity in our simulation and require that our model matches the chemical evolution of $\alpha$/Fe.

Using the techniques described in Section~\ref{sec:method},
a sample of stars enriched with the $r$-process in Eris are
shown in the top two panels of Figure~\ref{fig:EuFemix},
assuming that the only production channel for the
$r$-process is compact binary mergers.
The left panels of Figure~\ref{fig:EuFemix} show the results of our
simulation without metal diffusion and with no metallicity floor, whereas
the right panels show our fiducial model, which contains a simple
physically motivated prescription for the metal mixing, and a metallicity floor. In both
cases,  a large spread in Eu/Fe is produced at all times;
even after the Eu and Fe are diluted by metal-diffusion,
there exists a substantial scatter in Eu/Fe abundances.

\begin{figure*}
\plotone{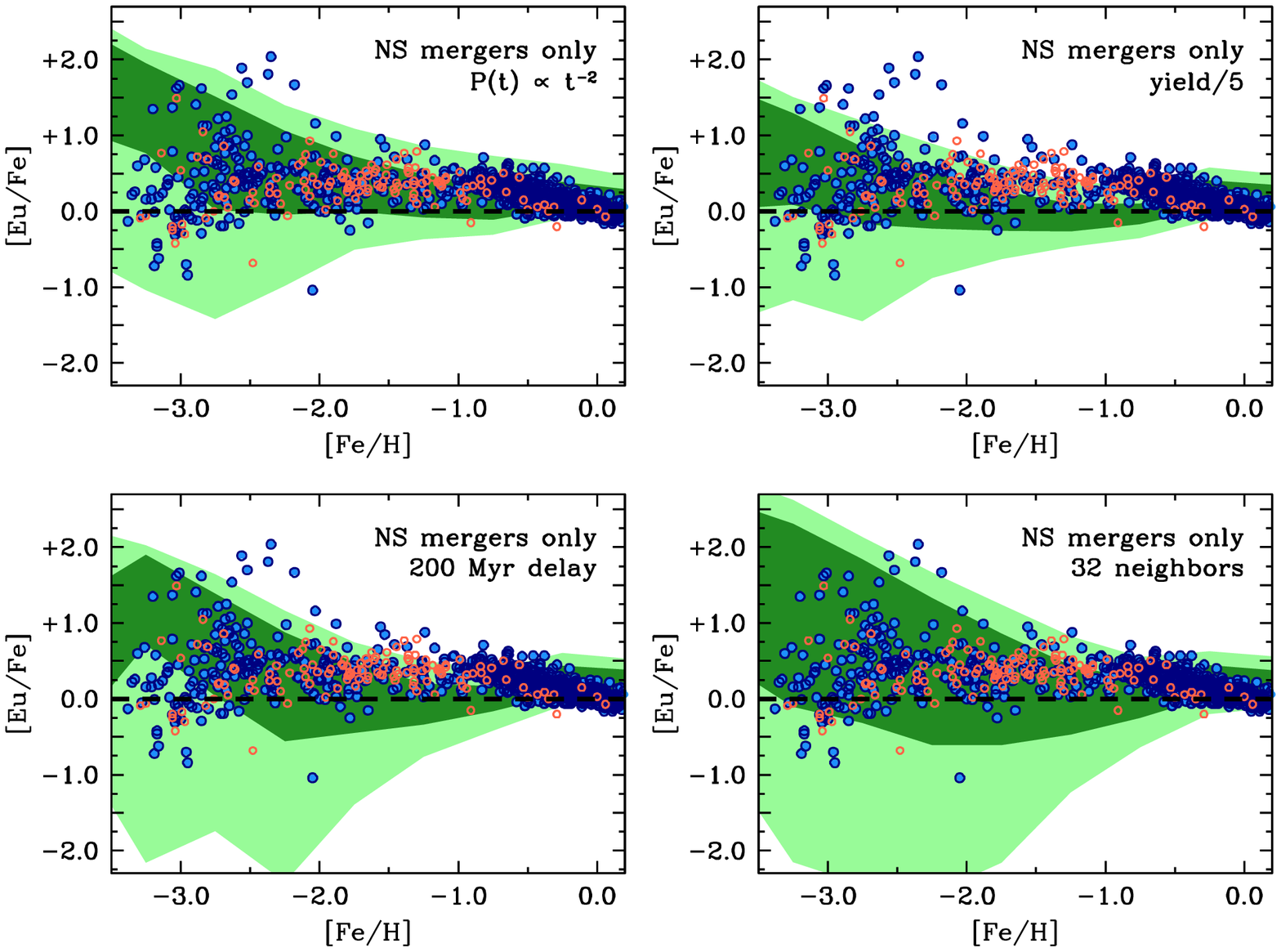}
\caption{A demonstration of how the simulated [Eu/Fe] contours depend on our model
parameters. We make the following changes to our fiducial model:
(1) change the DTD to be $P(t)\propto\ t^{-2}$ (top left panel);
(2) increase the merger rate and reduce the yield per event
by a factor of 5 (top right panel);
(3) double the merger time delay to $t_{\rm cut}=200~{\rm Myr}$ (bottom left panel); and
(4) less efficient mixing, by averaging 32 nearest neighbors when a star particle is formed (bottom right panel).
The dark and light green contours show the [Eu/Fe] distributions that
respectively enclose 68 and 95 per cent of Eris' star particles, with our mixing
prescription applied. The observational data are represented by blue symbols
for stars where $\alpha$, Fe, and Eu were all measured, or by orange symbols
for stars that only have Eu and Fe measured.
The dashed line in these panels represents the solar [Eu/Fe].}
\label{fig:summary}
\end{figure*}

We present the corresponding distributions
of the stellar [Eu/Fe] values with [Fe/H] metallicity
in Eris as the green contours in the
bottom panels of Figure~\ref{fig:EuFemix} (dark and light shades
enclose 68 and 95 per cent of Eris' star particles). The
observational data are drawn from the samples described
above, where blue symbols represent stars where $\alpha$, Fe,
and Eu were all measured, and orange symbols are for stars
where only Eu and Fe (but not $\alpha$) were reported. The most
striking feature of these panels is that NS mergers are able to
produce a significant scatter, even at low metallicity and early
times, which agrees with the observational data.
The bottom-right panel of Fig.~\ref{fig:EuFemix} illustrates our
fiducial model, which includes a prescription for mixing and
a metallicity floor. Our model simultaneously provides an acceptable
fit to both the [$\alpha$/Fe] distribution and the [Eu/Fe] distribution
for all metallicities.
Unfortunately, it is difficult to draw a direct comparison between the \textit{distribution}
of Eu/Fe values of the simulated and observed stars at low metallicity, since there
are a non-negligible number of stars where Eu has not yet been detected
\citep[e.g.][]{Bar05,Francois07,Cohen08,Roederer10,Roe14}.
Nevertheless, our models suggest that the highest Eu/Fe ratios are
produced at earlier times and at lower [Fe/H] metallicity.

\section{Discussion}

The density of free neutrons required for  $r$-process synthesis  points to an  explosive environment \citep{Bu1957}. \SNII\ have long been prime suspects although not enough is known about the detailed physics such as the explosion mechanism, the role of neutrino interactions  and the equation of state in order to create realistic models that actually synthesize $r$-process material \citep[e.g.][]{Roberts10}.
NS mergers offer a possible alternative to SNe as the primary  $r$-process site \citep{Lat1977}. The conditions estimated to characterize the decompressed ejecta from compact binary mergers  
is  compatible with the production of an $r$-process abundance pattern generally consistent with solar system matter, and in particular the third $r$-process peak.
The most recent numerical studies of matter ejected in such relativistic  mergers  shows specifically that the $r$-process heavy nuclei  are produced in solar proportion \citep{Roberts11,Bauswein13,Grossman13}. The differences between these two mechanisms should be discernible in the enrichment pattern of $r$-process elements in the Milky Way.

Since the chemical evolution of any galaxy is intrinsically tied to
its SFH, it is crucial to adopt a galaxy model that is
characteristic of the Milky Way. Eris is
a close analog of the Milky Way at $z=0$,
making it an ideal laboratory to study the chemical
evolution of $r$-process elements in the Galaxy. 
When we consider a simple prescription for chemical mixing, overall we
find an acceptable agreement between observations and Eris
for the relative production of $\alpha$-capture and Fe-peak elements.
By including a metallicity floor, we find that the [$\alpha$/Fe] abundance measured
for Eris is enhanced when [Fe/H]~$\lesssim-1.0$, with very little scatter, in agreement
with observations.

\begin{figure}
\plotone{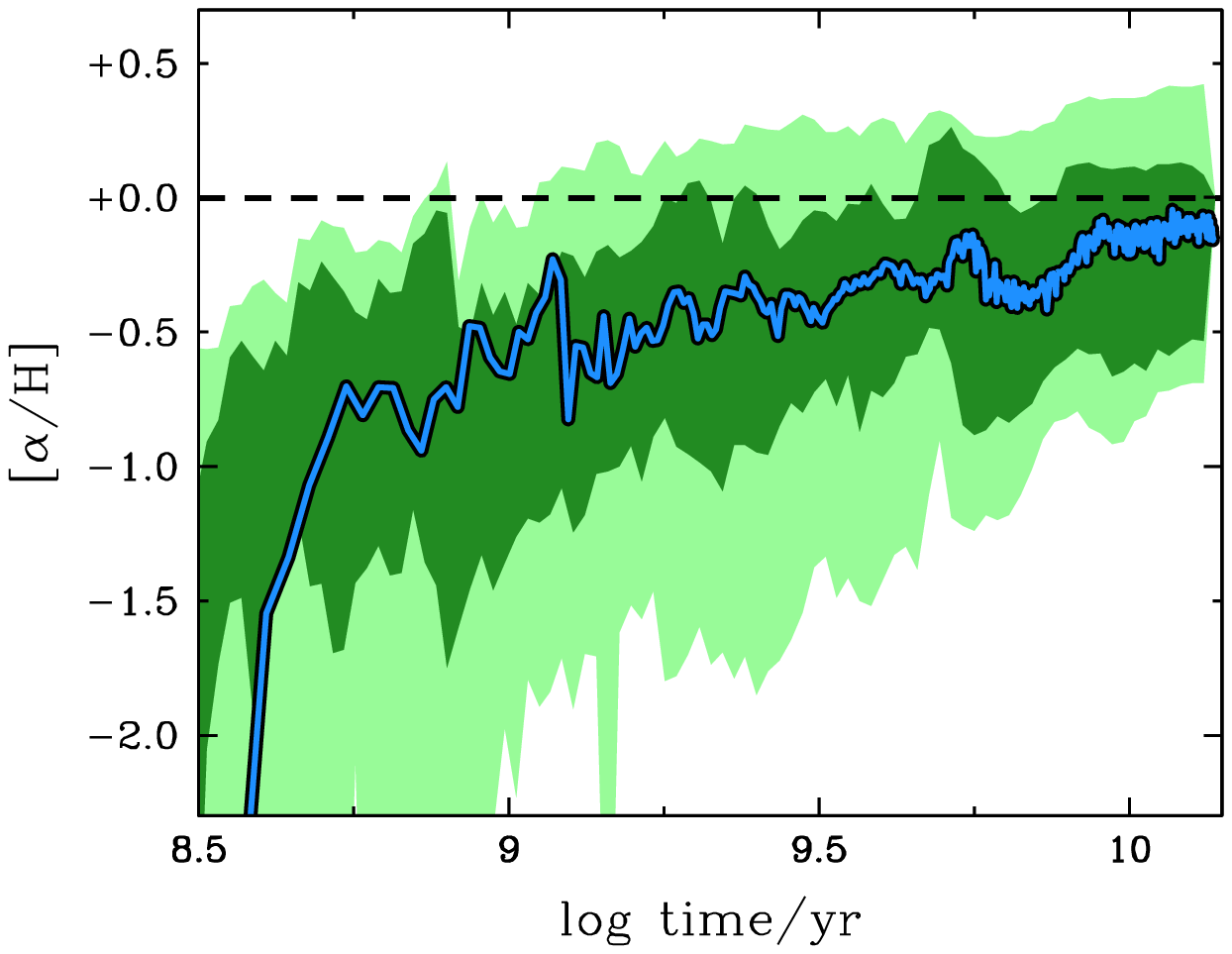}
\plotone{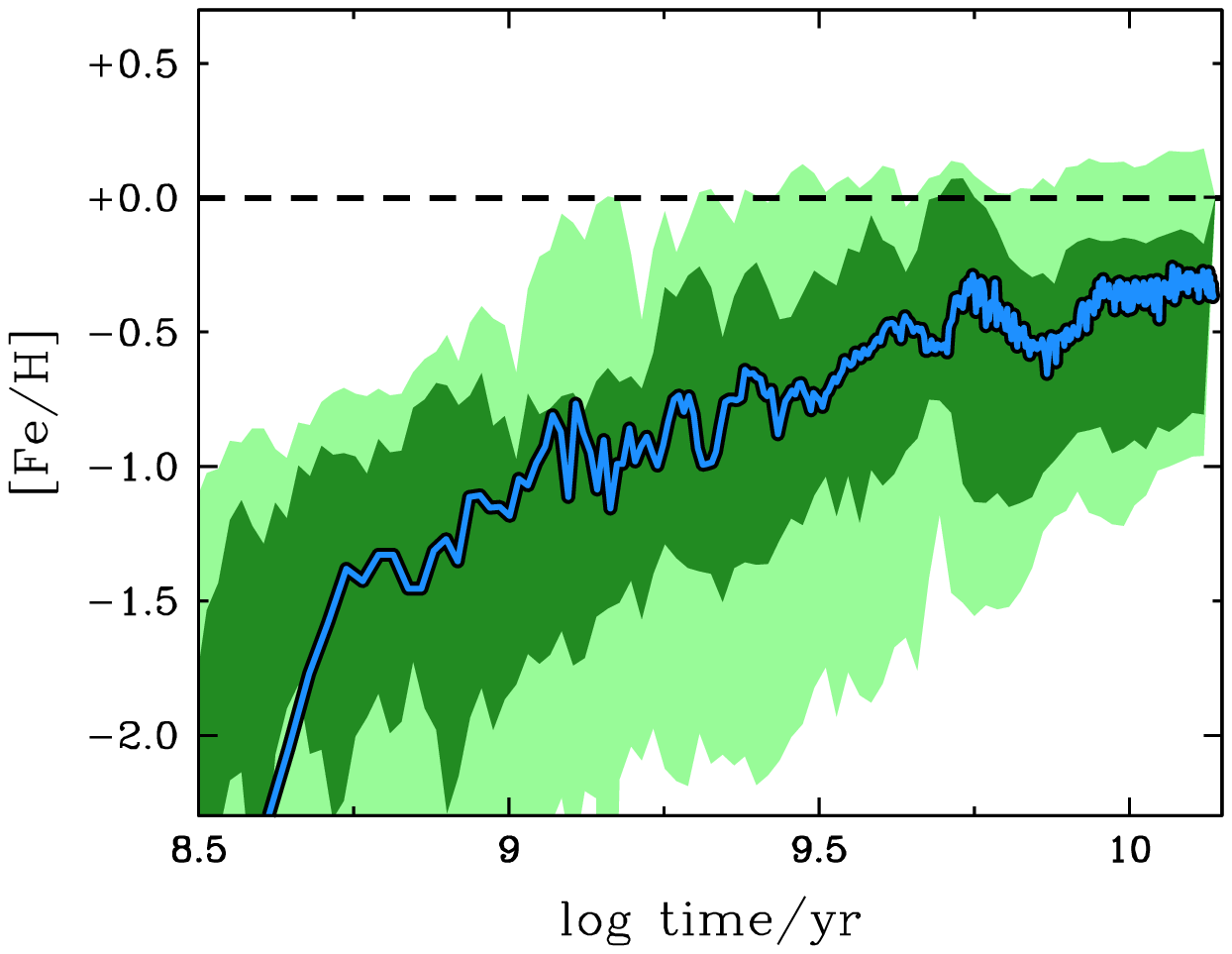}
\plotone{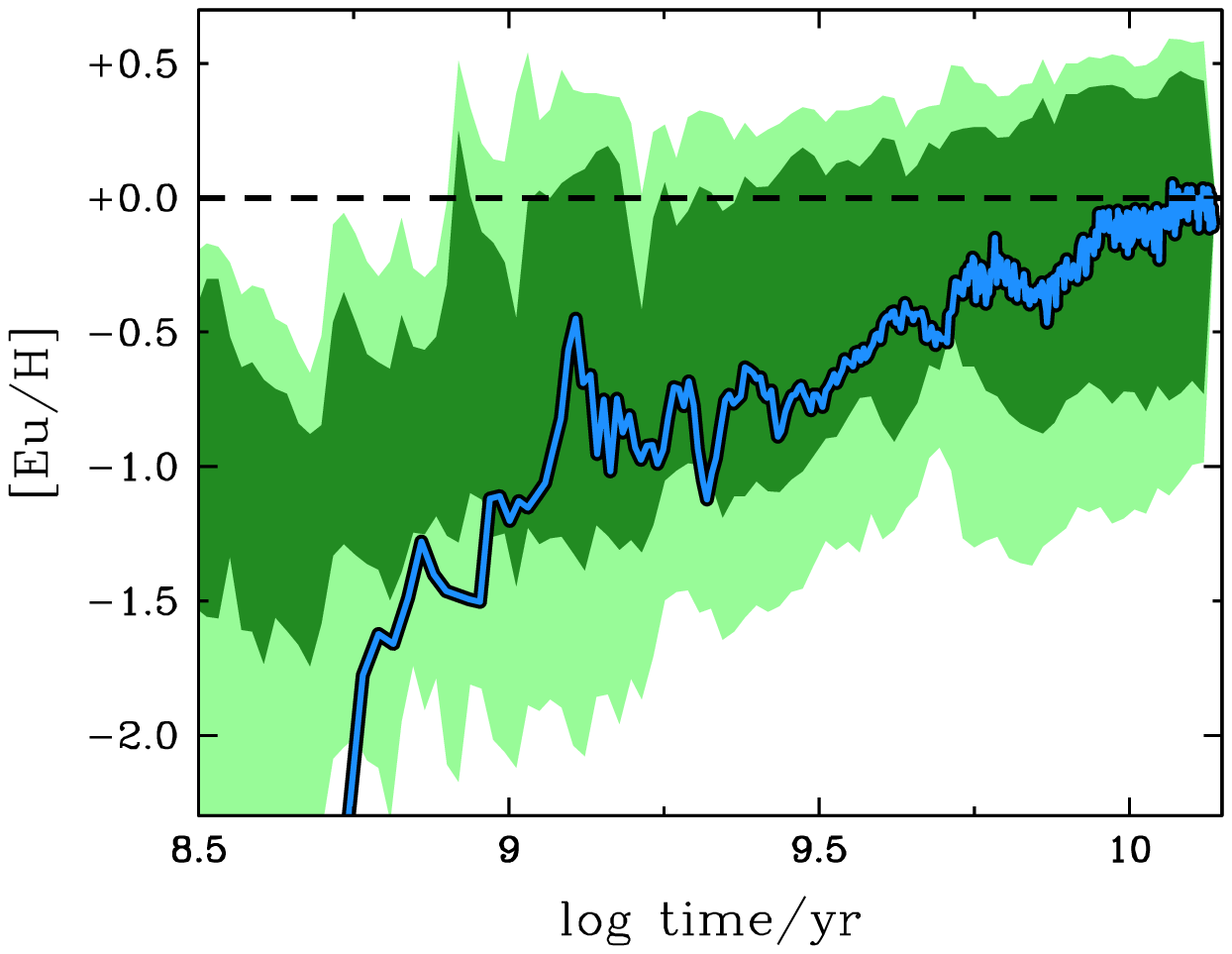}
\caption{The build-up of O, Fe, and Eu in Eris after introducing a metallicity floor of
$Z\sim10^{-4}Z_{\odot}$ and a simple mixing prescription. The dark and light green
regions show the 68 and 95 per cent contours that represent the metallicity distribution
of stars that are born at a given time. The blue curve represents the average metallicity
of the cold gas that is about to form a new generation of stars, and is therefore equivalent
to a one dimensional chemical evolution model with complete chemical mixing. The significant metallicity spread of the cold gas,
particularly for the high levels of Eu at early times, is not captured by a completely-mixed chemical
evolution model. The horizontal dashed line shows the solar level.}
\label{fig:timeevolution}
\end{figure}

\begin{figure*}
\plotone{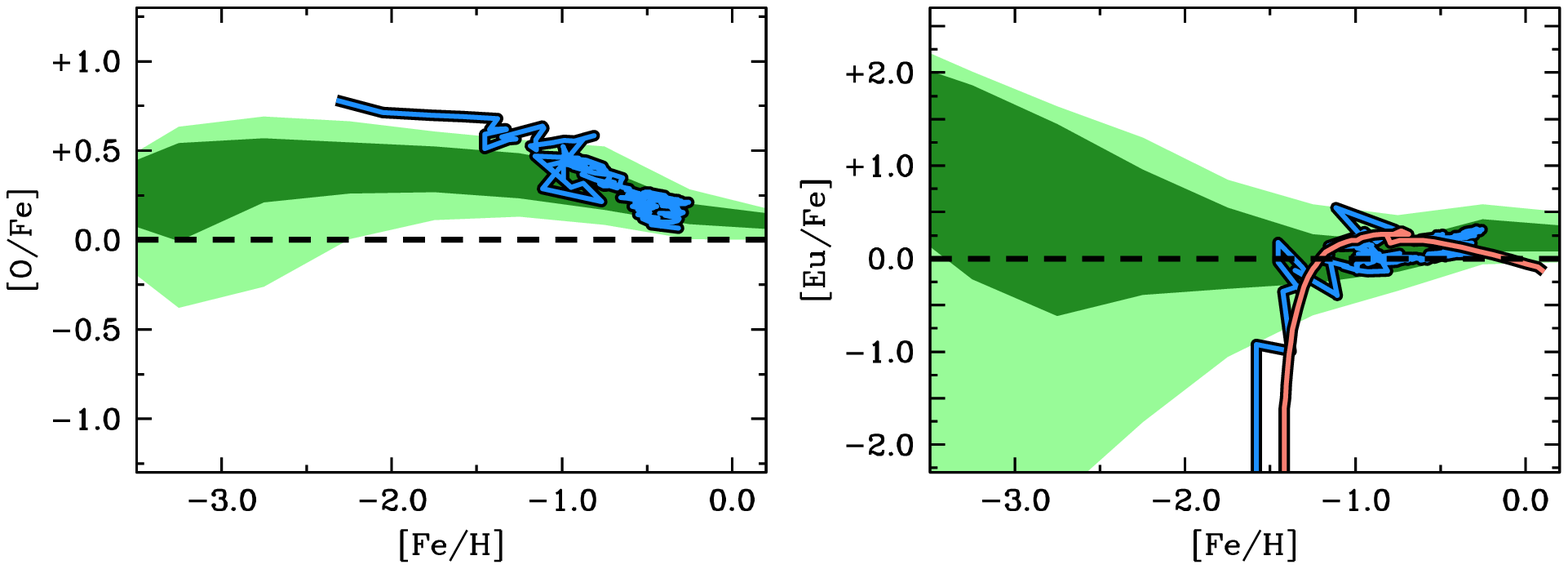}
\caption{The chemical evolution of O and Eu relative to Fe for our fiducial model. The dark
and light green contours enclose 68 and 95 per cent of Eris' star particles. The blue curve
shows the average abundance of the cold gas in Eris at each timestep that is about to
form a new generation of stars (i.e. the equivalent of a one-dimensional chemical evolution
model). Note the strong dissimilarity between the Eris results (green contours) and the completely-mixed Eris model (blue curves).
The red curve illustrates model {\it Mod1NS'} from \citet{Mat14}; this chemical evolution
model shares the closest similarity to our NS merger prescription. The horizontal dashed
black line represents the solar level of [Eu/Fe].}
\vspace*{\baselineskip}
\label{fig:wellmixed}
\end{figure*}

Previous studies that have investigated the chemical evolution of the $r$-process
with NS mergers have suggested that the merger timescale needs to be
relatively short ($\sim1$~Myr) in order for the $r$-process to be
borne into stars with metallicities [Fe/H]~$\lesssim-3.0$ \citep{Arg04,Mat14}.
Our analysis, on the other hand, suggests that the chemical enrichment
from NS mergers can be incorporated into stars of low [Fe/H] and at early
epochs even with a delay time of $\sim100$~Myr. Thus, the
relatively longer merger timescale argument no longer appears to
be a barrier for producing $r$-process enhanced stars at  low metallicity.

\subsection{Dependence on Model Parameters}
In our investigation, we have found that compact binary mergers
are able to successfully produce a large dispersion in the [Eu/Fe] ratios at
low metallicity, which is diluted at higher metallicity due to
chemical mixing. In general, the large $r$-process yield
and the rarity of NS mergers can produce a highly varied enrichment of
the $r$-process. We have also explored extensions to our
fiducial model including:
(1) a DTD with $P(t)\propto t^{-2}$ (top left panel of Fig.~\ref{fig:summary});
(2) increasing the frequency of mergers by a factor of five, whilst reducing the yield
per event by the same amount (top right panel of Fig.~\ref{fig:summary});
(3) doubling the initial time delay for mergers after a burst of star formation,
to $t_{\rm cut}=200~{\rm Myr}$ (bottom left panel of Fig.~\ref{fig:summary}); and
(4) adopting a less efficient mixing algorithm (bottom right panel of Fig.~\ref{fig:summary}).
A general result of decreasing the power-law index of the DTD,
is that more events will occur at earlier times (i.e. shortly
after the 100 Myr delay that we have imposed). The elevated number
of events at early times acts to modestly reduce the dispersion in [Eu/Fe]
at low metallicity. Similarly, by decreasing the Eu yield per event, the
number of events are globally increased in order to match the observed
Eu/O ratio in the Sun. The increased number of events also acts to slightly
reduce the dispersion in [Eu/Fe] at all metallicities.
When we change the merger delay time to be $t_{\rm cut}=200~{\rm Myr}$, the
evolution of [Eu/Fe] is qualitatively very similar to our fiducial model for [Fe/H]~$\gtrsim-2.0$.
However, at early times and hence lower metallicities ([Fe/H]~$\lesssim-2.0$),
there are subtle differences between this model extension and the fiducial model,
which are largely because there are fewer lower metallicity stars being enriched
with Eu. Finally, if we reduce the efficiency of mixing relative to the fiducial model,
by averaging the metallicity of the 32 nearest gas particles when a star
particle is born, our model provides a poorer description of the [Eu/Fe]
chemical evolution, particularly for the low [Eu/Fe] stars\footnote{We also
note that the [$\alpha$/Fe] chemical evolution is poorly matched with this
less efficient mixing prescription.}.

We therefore conclude that modest variations of the model parameters
for the injection history of the $r$-process, do not significantly alter
our results -- Eris displays a general correspondence with the chemical
evolution of [Eu/Fe] observed in the Milky Way. The choice of mixing
does not significantly affect the [Eu/Fe] distribution of the strongly
Eu-enhanced stars, where [Eu/Fe]$\gtrsim0.0$, and both models
provide an acceptable fit to the data in this regime. The choice of
mixing does, however, affect the distribution of the lowest [Eu/Fe] stars.
This regime is also the most poorly constrained by observations, since
there are many stars where the Eu abundance has not been measured
\citep[e.g.][]{Bar05,Francois07,Cohen08,Roederer10,Roe14}.
The level of mixing is therefore best determined
by using the abundance distribution of other elements, such as the
[$\alpha$/Fe] ratio as implemented herein. We encourage future
studies that consider a wider range of chemical elements to obtain
a better handle on the mixing process.

\subsection{Comparison with 1D Models} 
The chemical inhomogeneity of the interstellar medium is a key aspect
of chemical evolution that can be better addressed in numerical simulations. We
now explore the effect of chemical inhomogeneities and metal-mixing
in more detail. In particular, we consider a scenario that is similar to a 1D chemical
evolution model where the metal content of all star-forming gas (i.e. gas that turns
into stars at the next simulation timestep) is completely mixed before
forming a new generation of stars. This assumption has been used
in many analytic and semi-analytic calculations of chemical evolution.
Such models provide a greater degree of physical intuition, but
are unable to capture all of the relevant physical processes that are
important for galactic chemical evolution; for example, these models
are not typically set in a cosmological setting. The results of our
1D experiment are shown in Figure~\ref{fig:timeevolution},
where the dark and light contours display the 68 and 95 per cent
contours enclosing the chemical evolution of our fiducial model,
and the single blue line illustrates the result for our 1D experiment.
At late times, the centroid of both distributions agree very well,
although there is clearly a very large dispersion in all of the
element abundances that are considered in this work. However
at early times, and hence lower metallicities, our completely-mixed
chemical evolution model provides a very poor similarity to
the fiducial model. Moreover, the 1D model is unable
to keep track of the \textit{dispersion} at early times (see also
\citealt{Arg04}), which is particularly important for Eu. 

Figure~\ref{fig:wellmixed} further illustrates the shortcomings of
1D chemical evolution models by  plotting  the  evolution of O and Eu relative to Fe. 
It is evident that in the 1D models, stars with metallicities [Fe/H]~$\lesssim-3.0$  
cannot be formed after $\gtrsim$ 100 Myrs and that most of
the star-forming material in the Milky Way has reached  a level of [Fe/H]~$\sim - 3.0$   in the
$z\sim 15-20$  range. The ubiquity of Eu  in stars with  [Fe/H]~$\lesssim-3.0$ in our cosmological simulations
indicates that metal diffusion is not   widespread  from the sites of NS mergers until after a few Gyrs.
This heavy metal flow, mixed  and sheared by gas motions, remains highly anisotropic and is responsible for generating  the larger-scale Eu abundance variations  pervading low [Fe/H] stars.
The red curve in Figure~\ref{fig:wellmixed} contrasts the model {\it Mod1NS'} from \citet{Mat14} with our cosmological chemical evolution. Their  {\it Mod1NS'}
model shares the closest similarity to our NS merger prescription and as such provides the clearest model comparison.  It is clear that 1D chemical evolution models may misrepresent the true chemical
evolution of a galaxy, particularly at early times when metal mixing
is not yet effective. Future studies of
detailed chemical evolution in a cosmological setting
\citep[e.g.][]{KobNak11,MaiTesCoo15} are thus encouraged
to study the properties of metal-poor gas and stars.

\begin{figure}
\plotone{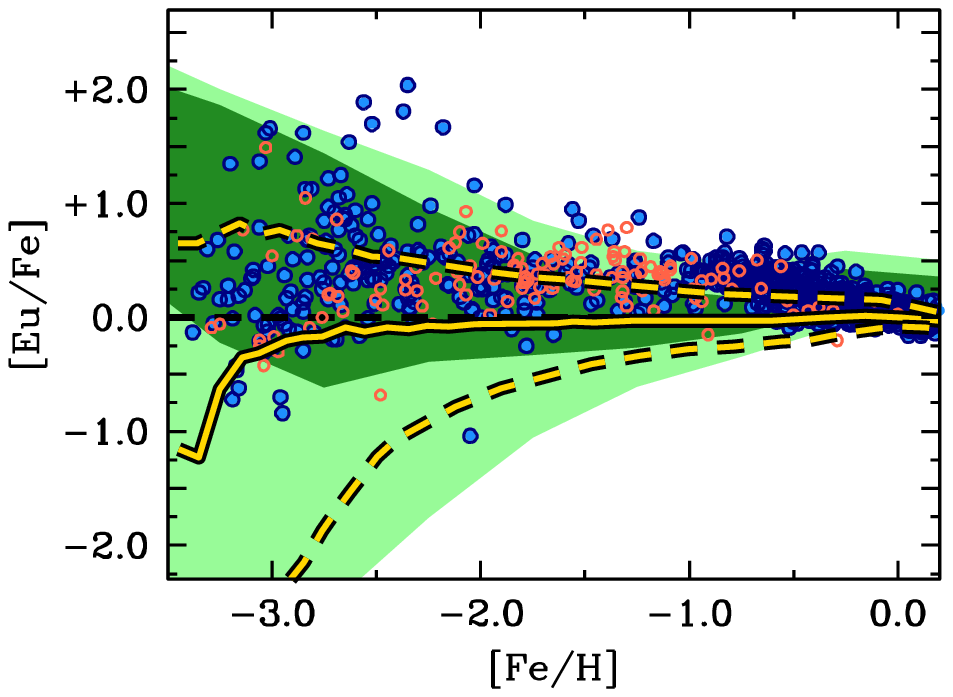}
\caption{Comparing our fiducial model for the chemical evolution of Eu relative to Fe, with the recent
simulations by \citet{vdVoort15}. The dark and light green contours enclose 68 and 95 per cent of
Eris' star particles. The solid yellow curve shows the distribution centroid of the fiducial model considered
by \citet{vdVoort15}, and the dashed yellow contours enclose 68 per cent of their star particles. The black
dashed line represents the solar level of [Eu/Fe].}
\label{fig:vdv}
\end{figure}

\subsection{Comparison with van de Voort et al. (2015)}
After the submission of this paper we became aware of a recent preprint \citep{vdVoort15}, in which the authors simultaneously reached similar conclusions as presented herein regarding the role of NS mergers in Galactic $r$-process enrichment using the cosmological zoom-in FIRE simulation \citep{Hopkins14}. From the parameter study by \citet{vdVoort15}, they also note that the abundance ratio and scatter of the $r$-process/Fe ratio at low metallicity may be sensitive to uncertainties in modelling turbulent mixing of the ISM in their simulation. In Fig.~\ref{fig:vdv}, we compare the chemical evolution of Eu/Fe for their fiducial model relative to ours. Both simulations appear to track the solar abundance of Eu/Fe for [Fe/H]~$\gtrsim-2.0$, and exhibit a similar level of scatter. Although both simulations exhibit a large dispersion of [Eu/Fe] values for [Fe/H]~$\lesssim-2.0$, the centroid of the two simulations shows an opposite evolution, 
which likely reflects the different model choices of Eu injection sites, numerical resolution, star formation history and feedback prescriptions between the two simulations.

Like all chemical evolution models, our conclusions may
be sensitive to the assumptions of our model, such as the
implementation of metal diffusion as well as the other
numerical choices outlined in Section~\ref{sec:erissim}.
It is nevertheless encouraging that our chemodynamical simulation with a simple, first order
prescription of sub-grid metal mixing provides an acceptable
agreement to the observed chemical evolution of
[$\alpha$/Fe]. Future cosmological simulations that
follow the chemical evolution of elements which are
produced on different timescales and from a larger
variety of production sites may help to elucidate some
of the finer details of chemical mixing. 

Our results demonstrate the importance of tracing both the dynamics and chemical evolution self-consistently in studying the origins and the evolution of $r-$process elements. In this regard, chemodynamic simulations are advantageous in modeling the chemical inhomogeneity in the ISM, and are highly complementary to computationally inexpensive analytic calculations that provide a somewhat higher level of physical intuition. Future analytic and semi-analytic calculations that include a prescription of
the effects of chemical inhomogeneity and mixing may
provide a very useful insight into galactic chemical
evolution, especially for trace elements, such as
the $r$-process.

Whilst there need not be a single $r$-process
production site, the current data favor a source that is either uncommon or whose
yield only sporadically produced the $r$-process, as argued by many other authors in the past.
Our study has found that NS mergers are a strong candidate as a dominant
production site, even at low metallicity. Future cosmological hydrodynamic
simulations with a more realistic mixing prescription, that follows a galaxy
with a similar chemical evolution and SFH to that experienced by the
Milky Way, are now required to investigate this problem in further detail.

\acknowledgments

We would like to thank L.~Bildstein and M.~Trenti for  helpful discussions,
and an anonymous referee for constructive comments. 
We gratefully acknowledge the hospitality of the Aspen Center for Physics and the DARK Cosmology Center while completing this work (NSF AST-1066293), and  support from
NSF (AST-1109447 and AST-0847563) and the David and Lucile Packard Foundation.
R.~J.~C. was partially supported by NSF grant AST-1109447
during this work, and is currently supported by NASA through
Hubble Fellowship grant HST-HF-51338.001-A, awarded by the
Space Telescope Science Institute, which is operated by the
Association of Universities for Research in Astronomy, Inc.,
for NASA, under contract NAS5- 26555.
P.M. acknowledges support by NSF grant OIA-112445329745 and  NASA grant NNX12AF87G.

\end{document}